\documentclass[aps,showpacs,prb,print,onecolumn,superscriptaddress]{revtex4}
\usepackage{mathrsfs}
\usepackage{amssymb}
\usepackage{amsmath}
\usepackage{bm}
\usepackage{graphics}
\usepackage{natbib}

\newcommand{\sss}[1]{{\scriptscriptstyle #1}}
\begin{document}
\title{Generalized transfer matrix theory on electronic transport through graphene waveguide}
\author{Haidong Li}
\address{National Laboratory of Superhard Materials, Department
of Physics, Jilin University, Changchun 130023, China}
\author{Lin Wang}
\address{National Laboratory of Superhard Materials, Department
of Physics, Jilin University, Changchun 130023, China}
\author{Zhihuan Lan}
\address{Department of Aviation Ordnance Engineering, Air Force
Aviation University, Changchun 130022, China}
\author{Yisong Zheng}\email[Correspondence author. Email: ]{zys@mail.jlu.edu.cn}
\address{National Laboratory of Superhard Materials, Department
of Physics, Jilin University, Changchun 130023, China}
\date{\today}

\begin{abstract}
In the effective mass approximation, electronic property in graphene
can be characterized by the relativistic Dirac equation. Within such
a continuum model we investigate the electronic transport through
graphene waveguides formed by connecting multiple segments of
armchair-edged graphene nanoribbons of different widths. By using
appropriate wavefunction connection conditions at the junction
interfaces, we generalize the conventional transfer matrix approach
to formulate the linear conductance of the graphene waveguide in
terms of the structure parameters and the incident electron energy.
In comparison with the tight-binding calculation, we find that the
generalized transfer matrix method works well in calculating the
conductance spectrum of a graphene waveguide even with a complicated
structure and relatively large size. The calculated conductance
spectrum indicates that the graphene waveguide exhibits a
well-defined insulating band around the Dirac point, even though all
the constituent ribbon segments are gapless. We attribute the
occurrence of the insulating band to the antiresonance effect which
is intimately associated with the edge states localized at the
shoulder regions of the junctions. Furthermore, such an insulating
band can be sensitively shifted by a gate voltage, which suggests a
device application of the graphene waveguide as an electric
nanoswitch.
\end{abstract}

 \maketitle

\bigskip

\section{Introduction}
The fabrication of graphene, a graphitic sheet of one-atom
thickness, is the first experimental realization of truly
two-dimensional crystal.\cite{refNovoselov1} Such a carbon material
presents many unusual electronic and transport properties, such as
the half-integer quantum Hall effect,
\cite{refzheng,refGusynin,refYb.Zhang,refNovoselov,refPeres} the
nonzero conductivity minimum at vanishing carrier
concentration,\cite{refNovoselov,refM.I.Katsnelson1,refJ.Tworzydlo,refAKGeim,refFMiao,refYW.Tan,refK.Ziegler,refK.Ziegler2}
the subtle weak
localization,\cite{refSuzuura,refS.V.Morozov,refA.F.Morpurgo,refKhveshchenko,refE.McCann,refXiaosongWu}
and the reflectionless transmission of the carrier through an
arbitrarily high
barrier.\cite{refKrekora,refM.I.Katsnelson,refV.V.Cheianov} From the
application point of view, graphene has very high mobility and shows
am-bipolar behavior.\cite{refS.Latil} More importantly, the planar
geometry of graphene makes it relatively easier to fabricate various
integrated nano-circuits. Therefore, graphene is regarded as a
perspective base for the post-silicon electronics.
\par
Motivated by possible device applications, electronic transport
through various graphene nano-structures were extensively studied
both experimentally and
theoretically.\cite{refD.A.Areshkin,refB.Obradovic,refM.Y.Han,refY.Ouyang,refA.Rycerz,refT.B.Martins,refhod,refL.Brey,refZ.p.Xu,refS.Hong,refT.C.Li}
Among these structures, graphene nanoribbon(GNR) is the basic
element to carry the current flow. Band structure calculation
indicates that the zigzag-edged GNR is always metallic while the
armchair-edged ones are either metallic or semiconducting, depending
on their width.\cite{refL.Brey} Recent experimental work has
confirmed the possibility of tunnelling the transport gaps of GNRs
by changing their widths.\cite{refM.Y.Han}
\par
A graphene junction can be formed by interconnecting two
semi-infinite GNRs with different widths. In such a graphene
nanostructure, a traveling carrier is scattered by the junction
interface. Thus, a finite junction conductance
appears.\cite{refZ.p.Xu,refS.Hong} Furthermore, when connecting
multiple segments of distinct GNRs in cascade manner, one can build
up graphene multiple junction(GMJ) structure. It was experimentally
demonstrated that the ballistic transport in graphene can be
retained over sub-micron scale.\cite{refYb.Zhang} Therefore, the
scattering of geometrical shape of a GMJ plays a dominant role in
determining its electronic transport property. Recently, the
conductance spectrum of a graphene single junction is studied in
details.\cite{refZ.p.Xu,refS.Hong} It was found that the presence of
the lattice vacancy can efficiently enhance the junction
conductance,\cite{refT.C.Li} because that a vacancy makes the
coupling between the electron states of the two GNRs at the junction
interface stronger. In contrast, the investigation on electronic
transport through GMJs and other graphene waveguides is still in its
infancy. However, one can reasonably expect that the graphene
waveguides possess many interesting electronic transport properties
controlled by their geometrical shapes. For instance, in contrast to
a graphene single junction, the GMJ has more junction interfaces.
Thus more electron partial waves, including the incident wave and
the multiple reflected waves from different interfaces, will take
part in the quantum interference. As a result, some transport
features observed in graphene single junction can be enhanced or
weakened, depending on the details of the geometrical shape of the
GMJ. Prior to the experimental realization to fabricate graphene
waveguides with high edge-order, theoretical prediction on the
geometrical shape dominated electronic transport properties of
typical graphene waveguides is highly desirable.
\par
So far two kinds of theoretical approaches, the tight-binding method
and the continuum model, have been frequently employed to study the
electronic and transport properties of graphene bulk material and
graphene
nanostructure.\cite{refN.M.R.Peres,refV.M.Pereira,refYMBlanter,refL.Brey1}
The tight-binding method is suitable to treat the electronic state
of graphene nanostructures with arbitrary shapes. Combined with the
Landauer-B\"{u}ttiker formula in discrete lattice
representation,\cite{refF.M,Datta} this method is convenient to
study the electronic transport through graphene nanostructures, with
various scattering mechanisms included naturally. However, as the
size of the graphene nanostructure under consideration increases,
the tight-binding method runs into an embarrassment since it needs
to treat some large matrices. Hence the calculation becomes rather
time-consuming. In contrast, such a problem does not occur in the
continuum model which is established based on the effective mass
approximation.\cite{refD.P.DiVincenzo,refT.Ando} The continuum model
has succeeded in describing the electronic and transport properties
of the bulk graphene. When applying it to the nanostructures with
relatively large sizes and regular shapes, some appropriate boundary
conditions are needed to solve the Dirac equation in this model.
\par
In the present work, we will investigate theoretically the
electronic transport through the GMJ structures. Such a graphene
waveguide usually has relatively large size. Thus, we will adopt the
continuum model in our theoretical treatment. To study the transport
properties we known that the transfer matrix approach is well
developed to describe the electronic transport through various
quantum waveguides made of conventional semiconductor
materials.\cite{refA.Weisshaar,refA.Weisshaar2,refW.D.Sheng,refJ.B.Xia}
However, when we attempt to apply this approach to graphene
waveguide, some generalization is necessary since the electron in
graphene obeys the relativistic Dirac equation, rather than the
schr\"{o}dinger equation for the conventional semiconductor
materials. This is just the subject of our present work. By using
our generalized transfer matrix method, we calculate the conductance
spectrum of the GMJs. We find that the GMJs always present a
well-defined insulating band around the Dirac point, no matter
whether the GNRs in the GMJ are metallic or semiconducting. We
analyze that the origin of the appearance of the insulating band is
the antiresonance effect produced by the edge states localized at
the lateral zigzag edges of the junction interface. Furthermore, we
also find the position of the insulating band can be sensitively
adjusted by exerting a gate voltage under the GMJ. According to such
a feature, we suggest that the GMJ structures can be considered as a
device prototype of a nanoswitch.
\par
The rest of this paper is organized as follows: In Sec.\ref{theory},
a self-contained theoretical framework to formulate the linear
conductance of the GMJ structures is elucidated, though it is, in
fact, a generalization of the conventional transfer matrix method.
At first, we establish the relationship between the linear
conductance and the scattering matrix in the GMJ. Then we develop
the transfer matrix approach to work out the scattering matrix,
which is realized by means of the wavefunction connection condition
and boundary condition at the junction interfaces. In
Sec.\ref{result}, the numerical result on the linear conductance
spectrum of some typical GMJs are shown. The antiresonance driven
insulating band shown in the calculated conductance spectrum is
discussed. Finally, the main results are summarized in
Sec.\ref{summary}.

\par

\section{theory\label{theory}}
\subsection{Scattering matrix and linear conductance}

The geometry and the honeycomb lattice structure of the GMJ under
our consideration is illustrated in Fig.1. The relevant structure
parameters are also explained in the same figure. Obviously, the
basic elements consisting of such a GMJ are armchair-edged GNRs of
different widths. The two semi-infinite ribbons at the left and
right sides serve as two leads for electronic transmission. The
other finite segments of GNRs are in between, constituting the
device region where the electronic tunneling is scattered. It was
previously shown that the continuum model can describe the
electronic property of an infinitely long armchair-edged GNR very
well.\cite{refL.Brey2,refL.Brey} The electronic eigen-state in such
a ribbon obeys the following Dirac equation subjected to a certain
boundary condition.
\begin{eqnarray}
H\psi=\gamma \begin{bmatrix}0 & -\bm{\hat{k}}_- & 0& 0\\
 -\bm{\hat{k}}_+& 0& 0 & 0\\
 0  & 0 & 0 &  \bm{\hat{k}}_+\\
 0  & 0 & \bm{\hat{k}}_-  & 0
\end{bmatrix}
\begin{bmatrix}\phi_A\\
 \phi_B\\
 \phi_{A}^{'} \\
 \phi_{B}^{'}
\end{bmatrix}
=E
\begin{bmatrix}\phi_A\\
 \phi_B\\
 \phi_{A}^{'} \\
 \phi_{B}^{'}
\end{bmatrix}, \label{dirac}
\end{eqnarray}
where $\widehat{\bm{k}}_{\pm}=-i\partial_x\pm \partial_y $,
$\gamma=\sqrt{3}ta_{0}/2$ is the so-called Fermi velocity and $t$
being the electron hopping energy between the neighbor lattice
points. We choose the Dirac point as the energy reference point and
in what follows we use the units $\hbar=t=a_{0}=1$. The electronic
eigen-state in the above equation can be analytically solved, which
is given by
\begin{eqnarray}
\bm u_{jsk_j}(x,y)=\begin{bmatrix}\phi_A\\
 \phi_B\\
 \phi_{A}^{'} \\
 \phi_{B}^{'}
\end{bmatrix}
=\frac{1}{2\sqrt{W}}\left(\begin{array}{cc}
e^{iq_{j}x}\\
-se^{i\theta_{j}}e^{iq_{j}x}\\
-e^{-iq_{j}x}\\
se^{i\theta_{j}}e^{-iq_{j}x}\end{array}\right)e^{ik_jy}.
\label{eigenstate}
\end{eqnarray}
And the corresponding eigen-energy is $E=s\gamma\sqrt{q_j^2+k_j^2}$.
$s=\pm 1$ denotes the conduction and valence bands, respectively. In
Eq.(\ref{eigenstate}) $W$ denotes the width of the ribbon and its
length in y direction has been normalized as unity. $\theta_j$ is
the angle between the wavevector $\bm{k}=({q_j,k_j})$ and the $x$
axis. The transverse wavevector component $q_j$ takes some discrete
values due to the quantum confinement.
\begin{eqnarray}
q_j=\frac{2j\pi}{W}+\frac{2\pi}{3},\;\; j=0,\pm 1, \pm 2 \cdots.
\end{eqnarray}
When the ribbon width takes some particular values: $W=3m/2$ with
the integer $m\geq 2$, such a wavevector can be alternatively
expressed as $q_j=j\pi/W$. Thus, $q_j=0$ is allowed, which implies a
zero bandgap between the conduction and valence bands. In such a
particular case the GNR presents a metallic behavior.
\par
In each segment of the GMJ as shown in Fig.1, the wavefunction of
the electronic state with energy $E$ can be expanded in terms of the
eigen-states of the armchair-edged GNR with the same width. To be
concrete, such a wavefunction expansion in $\lambda$th segment can
be expressed as
\begin{equation}
\psi^\lambda(x,y)=\sum_j[a^\lambda_ju^\lambda_{jsk_j}+b^\lambda_ju^\lambda_{js\bar{k}_j}]+
\sum_m[\alpha^\lambda_m\tilde{u}^\lambda_{msp_m}+\beta^\lambda_m\tilde{u}^\lambda_{ms\bar{p}_m}]
\label{wfi},
\end{equation}
where the wavevector $k_j(p_m)\geq 0$ and
$\bar{k}_j(\bar{p}_m)=-k_j(p_m)$. Evidently, the first two terms in
the right-hand side of the above equation stand for the propagating
modes along the positive and negative directions, respectively.
Apart from these propagating modes, we need to add the evanescent
modes in the above expansion. The corresponding eigen wavefunction
is given by
\begin{equation}
\tilde{u}_{msp_m}=\left(\begin{array}{cc}
e^{iq_{m}x}\\
-\chi_m e^{iq_{m}x}\\
-e^{-iq_{m}x}\\
\chi_m e^{-iq_{m}x}\end{array}\right)e^{p_my}, \label{imode}
\end{equation}
with
\begin{equation}
\chi_m=s\sqrt{\frac{|q_m+p_m|}{|q_m-p_m|}}\text{sgn}(m).
\end{equation}
\par
Of course, these evanescent modes are unphysical solutions in any
uniform ribbon with infinite length. However, they are indispensable
in the present GMJ structure to make the wavefunction continuity at
the junction interfaces. Noting that in the left and right leads we
should let $\beta^1=\alpha^{\sss{N}+1}=0$ to prevent the
wavefunction from diverging. In Eq.(\ref{wfi}) the relation between
the wavevectors and the incident electron energy $E$ is
\begin{equation}
E-V=s\gamma\sqrt{q_j^2+k_j^2}=s\gamma\sqrt{q_m^2-p_m^2},
\end{equation}
where $V$ represents a constant potential to mimic a possible gate
voltage exerted in this segment. For a given energy $E$ the number
of the propagating mode appearing in the expansion, denoted by
$J^\lambda_r$, is finite. And it can be readily determined by the
above dispersion relation. On the other hand, one can also see from
the above dispersion relation that the number of evanescent modes is
in principle infinite. We can actually add an appropriate number of
the evanescent modes in the expansion, $J^\lambda_e$, to guarantee
the convergence of the calculated results. Thus the total mode
number in $\lambda$th unit is specified,
$J^\lambda=J^\lambda_r+J^\lambda_e$. Such a quantity varies among
the segments, depending on their respective widths.
\par
The central task in our theory is to find the scattering matrix
which links the injected and the reflected wave amplitudes from the
two leads to the device region. Namely
\begin{eqnarray}
\left[\begin{array}{cc} \bm{b}^1\\\bm{a}^{\sss{N}}
\end{array}\right]=[S]\left[\begin{array}{cc}
\bm{a}^1\\\bm{b}^{\sss{N}}
\end{array}\right].\label{scattering}
\end{eqnarray}
Where $\bm{a}^1$ and $\bm{b}^1$($\bm{a}^\sss{N}$ and
$\bm{b}^\sss{N}$) are column matrices consisting of the expansion
coefficients in Eq.(\ref{wfi}), namely, the wave probability
amplitudes of all the propagating modes in the left(right) lead. To
determine the scattering matrix $[S]$ one should make use of the
wavefunction continuity at all junction interfaces. And the formal
theory is referred to as the transfer matrix method. But before
going into the details about the transfer matrix, we had better to
present a brief statement about the relationship between the linear
conductance and the scattering matrix, particularly suitable to the
GMJ structure.
\par
First of all, we define a pseudo time reversal operator
\begin{eqnarray}
\hat{T}=(\sigma_{z}\otimes I)\hat{C}=\begin{bmatrix}
0&0&1&0\\
0&0&0&1\\
1&0&0&0\\
0&1&0&0\\
\end{bmatrix}\hat{C}\label{timereverseal},
\end{eqnarray}
with $\hat{C}$ being the complex-conjugate operators. It is
commutable with the Dirac Hamiltonian given in Eq.(\ref{dirac}).
Such a commutation means that in any segment of the GMJ, besides
$\psi^\lambda$, $\hat{T}\psi^\lambda$ are also the possible
wavefunction with the same eigen-energy. To be specific,
$\hat{T}\psi^1$ and $\hat{T}\psi^\sss{N}$ are allowable
wavefunctions in the left and right leads. Owing to such an argument
the following relation holds true,
\begin{eqnarray}
\left[\begin{array}{cc} {\bm{a}^1}^*\\{\bm{b}^{\sss{N}}}^*
\end{array}\right]=[S]\left[\begin{array}{cc}
{\bm{b}^1}^*\\{\bm{a}^{\sss{N}}}^*
\end{array}\right].\label{scattering2}
\end{eqnarray}
By comparing it with Eq.(\ref{scattering}) we find that the
scattering matrix satisfies
\begin{eqnarray}
[S]^{*}=[S]^{-1}.\label{sm}
\end{eqnarray}
To calculate the probability currents in both leads, we obtain
\begin{equation}
J_{L}=\langle\Psi^1|\hat{v}_y|\Psi^1\rangle= \underset{j}\sum[\gamma
\sin\theta^1_{j}|a^1_{j}|^2-\gamma\sin\theta^1_{j}|b^1_{j}|^2],
\end{equation}
and
\begin{equation}
J_{R}=\langle\Psi^\sss{N}|\hat{v}_y|\Psi^\sss{N}\rangle=
\underset{j}\sum[\gamma\sin\theta^\sss{N}_{j}|a^\sss{N}_{j}|^2-\gamma\sin\theta^\sss{N}_{j}|b^\sss{N}_{j}|^2].
\end{equation}
In the above two equations the velocity operator is defined as
$\hat{v}_y=i[H,\hat{y}]$. The probability current conservation
requires $J_{L}=J_{R}$, namely
\begin{equation}
\underset{j}\sum \sin\theta^1_{j}|a^1_{j}|^2+ \underset{j}\sum
\sin\theta^\sss{N}_{j}|b^\sss{N}_j|^2 =\underset{j}\sum
\sin\theta^1_{j}|b^1_{j}|^2+\underset{j}\sum
\sin\theta^\sss{N}_{j}|a^\sss{N}_j|^2. \label{current}
\end{equation}
If we define a new scattering matrix $[S']=[v][S][v]^{-1}$ in which
the diagonal matrix $[v]$ is defined as
$[v]=\text{diag}[\{\sin^{1\over 2}\theta^1_j\},\{\sin^{1\over
2}\theta^\sss{N}_j\}]$, we can derive from Eq.(\ref{current}) that
the matrix $[S']$ is unitary, namely
\begin{equation}
 [S'][S']^\dag=1.
\end{equation}
From the above relation and Eq.(\ref{sm}) we can deduce an important
conclusion about the matrix $[S']$. That is the self-transposability
of it.
\begin{equation}
[S']=[S']^T.
\end{equation}
\par
Now suppose that an incident electron comes from the $l$th mode in
the left lead. Then it has certain probability to penetrate into the
$j$th mode of the right lead. The corresponding transmission
coefficient is defined as the ratio of the incident and the
penetrated probability current. By a straightforward calculation we
find that such a quantity is associated with the scattering matrix
$[S']$. It is given by
\begin{equation}
T_{l\rightarrow j}=\frac{J_{j}}{J_{l}}=|[S]_{jl}|^2\frac{\sin
\theta^\sss{N}_j}{\sin\theta^1_l}=|[S']_{jl}|^2.
\end{equation}
Conversely, for an electron incident from the $j$th mode in the
right lead to enter into the $l$th mode in the left lead, we can
work out the transmission coefficient in the same way.
\begin{equation}
T_{j\rightarrow
l}=\frac{J_{l}}{J_{j}}=|[S]_{lj}|^2\frac{\sin\theta^1_l}{\sin
\theta^\sss{N}_j}=|[S']_{lj}|^2.
\end{equation}
The self-transposability of $[S']$ guarantees the symmetry of the
transmission coefficients, i.e.
\begin{equation}
T_{l\rightarrow j}=T_{j\rightarrow l}=T_{lj}.
\end{equation}
By virtue of such a relation, when a small bias voltage is
established between the left and right leads, the net current can be
compactly expressed as
\begin{equation}
I={e\over h}\int \sum_{jl}T_{jl}[f_\sss{R}(E)-f_\sss{L}(E)]dE,
\end{equation}
where $f_\sss{L}(E)$ and $f_\sss{R}(E)$ are the Fermi distribution
functions in both leads. Consequently, from the above expression we
can readily extract an expression of the linear conductance. It is
given by
\begin{eqnarray}
\mathcal G=\frac{e^{2}}{h}\underset{jl}\sum
T_{jl}=\frac{e^{2}}{h}\underset{jl}\sum |[S']_{jl}|^2.\label{e43}
\end{eqnarray}
\subsection{scattering matrix and transfer matrix}
Next we need to work out the scattering matrix with the help of the
connection condition of wavefunctions at the junction interfaces. To
illustrate our derivation, we consider the wavefunction continuity
at an arbitrary interface, say the $\lambda$th one, as shown in
Fig.1(b). It is expressed as
\begin{equation}
\psi^\lambda_\sss{A}(x,y^\lambda)=\psi^{\lambda+1}_\sss{A}(x,y^\lambda),
\;\; 0\leq x <W_{\lambda+1}, \label{continuity}
\end{equation}
and
\begin{equation}
\psi^\lambda_\sss{B}(x,y'^\lambda)=\psi^{\lambda+1}_\sss{B}(x,y'^\lambda),
\;\; 0\leq x <W_{\lambda+1}, \label{continuityb}
\end{equation}
where $\psi^\lambda_\sss{A}$($\psi^\lambda_\sss{B}$) is a
two-component spinor, representing the wavefunction component
belonging to $A(B)$ atoms in $\lambda$th segment. Noting that
$y'^\lambda=y^\lambda-(\sqrt{3})^{-1}$ in the above equations, it
arises from the spatial difference between the adjacent $A$ and $B$
atoms. However, such a trivial difference can be safely ignored
since in the continuum model which is particularly appropriate to
describe the large size structure, the envelop wavefunction in the
preceding equations is slow-varying in lattice constant scale. At
the shoulder region of the $\lambda$th junction interface, the
wavefunction component corresponding to the truncated $A$ atoms
should vanish. It yields a boundary condition as below
\begin{equation}
\psi^\lambda_\sss{A}(x,y^\lambda)=0,\;\; W_{\lambda+1}\leq x\leq
W_\lambda. \label{shoulder}
\end{equation}
At other junction interfaces there are analogous wavefunction
connection and boundary conditions. From these conditions we can
derive a relation about the expansion coefficients in the adjacent
segments. First of all, for convenience we adopt the following Dirac
notations to denote the respective components of the two
sublattices($A$ and $B$) in the GNR eigen-state. For $A$ atom it is
denoted as
\begin{equation} |\lambda A s^\lambda
lk_l^\lambda\rangle= {1\over
\sqrt{2W_\lambda}}\left(\begin{array}{c} e^{iq_l^\lambda
x}\\-e^{-iq_l^\lambda x}
\end{array}\right)e^{ik^\lambda_ly},
\end{equation}
and the one for $B$ atom is
\begin{equation}
|\lambda B s^\lambda lk_l^\lambda\rangle=-s^\lambda
e^{i\theta^\lambda_l}|\lambda A s^\lambda lk_l^\lambda\rangle.
\end{equation}
Such a separated expression is equivalent to the eigen wavefunction
given by Eq.(\ref{eigenstate}) except for the trivial change of the
normalization constant. For the evanescent modes allowed in the
junction structure, the eigen wavefunction given by Eq.(\ref{imode})
can also be denoted in a separated form.
\begin{equation}
|\lambda A s^\lambda mp_m^\lambda\rangle= {1\over
\sqrt{2W_\lambda}}\left(\begin{array}{c} e^{iq_m^\lambda
x}\\-e^{-iq_m^\lambda x} \end{array}\right)e^{p^\lambda_my},
\end{equation}
and
\begin{equation}
|\lambda B s^\lambda mp_m^\lambda\rangle=-\chi_m |\lambda A
s^\lambda mp_m^\lambda\rangle.
\end{equation}
Noting that to normalize these evanescent eigenmodes is not needed
because it does not affect the final form of the scattering matrix.
With these notations the wavefunction connection condition given by
Eq.(\ref{continuity}) can be rewritten in an expanded form.
\begin{eqnarray}
\sum_l[a^\lambda_l|\lambda A s^\lambda lk_l^\lambda\rangle+&
&\hspace{-4mm}b^\lambda_l|\lambda A s^\lambda
l\bar{k}_l^\lambda\rangle]+ \sum_m[\alpha^\lambda_m|\lambda A
s^\lambda mp_m^\lambda\rangle+\beta^\lambda_m|\lambda A s^\lambda
m\bar{p}_m^\lambda\rangle]=\nonumber\\
 & &\sum_j[a^{\lambda+1}_j|\lambda+1 A s^{\lambda+1}
jk_j^{\lambda+1}\rangle+b^{\lambda+1}_j|\lambda+1 A s^{\lambda+1}
j\bar{k}_j^{\lambda+1}\rangle]+ \label{link}\\ \nonumber & &
\sum_n[\alpha^{\lambda+1}_n|{\lambda+1} A s^{\lambda+1}
np_n^{\lambda+1}\rangle+\beta^{\lambda+1}_n|{\lambda+1} A
s^{\lambda+1} n\bar{p}_n^{\lambda+1}\rangle].
\end{eqnarray}
In order to find a relation between these expansion coefficients, we
act on the eigenkets in both sides of the above equation by the
eigenbra $\langle\lambda A s^\lambda lk_l^\lambda|$. Special
attention should be paid to the issue that the integration over $x$
involved in the inner product calculation should be restricted in
the narrow region of the junction, namely, $[0,W_{\lambda+1}]$ as
labeled in Fig.1. However, considering the boundary condition given
by Eq.(\ref{shoulder}), such integrations in the left-hand side of
Eq.(\ref{link}) can be safely extended to total interface, i.e.
$[0,W_\lambda]$. Following such a rule, we obtain
\begin{eqnarray}
a^\lambda_l+e^{-2ik_l^\lambda y^\lambda}b_l^\lambda
&=&\sum_j[\langle \lambda As^\lambda
lk_l^\lambda|\lambda+1As^{\lambda+1}jk_j^{\lambda+1}\rangle
a_j^{\lambda+1}+\langle \lambda As^\lambda
lk_l^\lambda|\lambda+1As^{\lambda+1}j\bar{k}_j^{\lambda+1}\rangle
b_j^{\lambda+1}] \label{cs1} \\ \nonumber & &+\sum_n[\langle \lambda
As^\lambda
lk_l^\lambda|\lambda+1As^{\lambda+1}np_n^{\lambda+1}\rangle
\alpha_n^{\lambda+1}+\langle \lambda As^\lambda
lk_l^\lambda|\lambda+1As^{\lambda+1}n\bar{p}_n^{\lambda+1}\rangle
\beta_n^{\lambda+1}].
\end{eqnarray}
Instead of $\langle\lambda A s^\lambda lk_l^\lambda|$, using the
eigenbra $\langle\lambda A s^\lambda mp_m^\lambda|$ to take the
inner product we have
\begin{eqnarray}
 e^{2p_m^\lambda y^\lambda}\alpha^\lambda_m+\beta_m^\lambda
&=&\sum_j[\langle \lambda As^\lambda
mp_m^\lambda|\lambda+1As^{\lambda+1}jk_j^{\lambda+1}\rangle
a_j^{\lambda+1}+\langle \lambda As^\lambda
mp_m^\lambda|\lambda+1As^{\lambda+1}j\bar{k}_j^{\lambda+1}\rangle
b_j^{\lambda+1}]\\ \nonumber & &+\sum_n[\langle \lambda As^\lambda
mp_m^\lambda|\lambda+1As^{\lambda+1}np_n^{\lambda+1}\rangle
\alpha_n^{\lambda+1}+\langle \lambda As^\lambda
mp_m^\lambda|\lambda+1As^{\lambda+1}n\bar{p}_n^{\lambda+1}\rangle
\beta_n^{\lambda+1}].\label{cs2}
\end{eqnarray}
The characteristic of the two above equations is that only one
specific mode in $\lambda$th segment(labeling as $l$ or $m$) is
left. And the corresponding coefficients is expressed as a function
of all the coefficients in $(\lambda+1)$th segment. In addition, if
we apply the above procedure to the $(\lambda-1)$th interface, we
can obtain two analogous equations.
\begin{eqnarray}
a^\lambda_l+e^{-2i\theta_l^\lambda}e^{-2ik_l^\lambda
y^{\lambda-1}}b_l^\lambda &=&\sum_j[\langle \lambda Bs^\lambda
lk_l^\lambda|\lambda-1Bs^{\lambda-1}jk_j^{\lambda-1}\rangle
a_j^{\lambda-1}+\langle \lambda Bs^\lambda
lk_l^\lambda|\lambda-1Bs^{\lambda-1}j\bar{k}_j^{\lambda-1}\rangle
b_j^{\lambda-1}]\label{cs3}\\ \nonumber & &+\sum_n[\langle \lambda
Bs^\lambda
lk_l^\lambda|\lambda-1Bs^{\lambda-1}np_n^{\lambda-1}\rangle
\alpha_n^{\lambda-1}+\langle \lambda Bs^\lambda
lk_l^\lambda|\lambda-1Bs^{\lambda-1}n\bar{p}_n^{\lambda-1}\rangle
\beta_n^{\lambda-1}],
\end{eqnarray}
and
\begin{eqnarray}
\chi_m^2 e^{2p_m^\lambda
y^{\lambda-1}}\alpha^\lambda_m+\beta_m^\lambda &=&\sum_j[\langle
\lambda Bs^\lambda
mp_m^\lambda|\lambda-1Bs^{\lambda-1}jk_j^{\lambda-1}\rangle
a_j^{\lambda-1}+\langle \lambda Bs^\lambda
mp_m^\lambda|\lambda-1Bs^{\lambda-1}j\bar{k}_j^{\lambda-1}\rangle
b_j^{\lambda-1}]\label{cs4}\\\nonumber & &+\sum_n[\langle \lambda
Bs^\lambda
mp_m^\lambda|\lambda-1Bs^{\lambda-1}np_n^{\lambda-1}\rangle
\alpha_n^{\lambda-1}+\langle \lambda Bs^\lambda
mp_m^\lambda|\lambda-1Bs^{\lambda-1}n\bar{p}_n^{\lambda-1}\rangle
\beta_n^{\lambda-1}].
\end{eqnarray}
Combining Eqs.(\ref{cs1}-\ref{cs4}) and by a straightforward
derivation, it is possible for us to establish an iteration relation
between the expansion coefficients of the adjacent segments. It is
written in a matrix form as
\begin{equation}
[C^\lambda]=[M^{\lambda}_{\lambda-1}][C^{\lambda-1}]+
[M^{\lambda}_{\lambda+1}][C^{\lambda+1}], \;\; 2\leq\lambda\leq
{N-1}, \label{matrix}
\end{equation}
where $[C^\lambda]$ is defined as a column matrix consisting of all
expansion coefficients in $\lambda$th segment, namely,
$[C^\lambda]=[a^\lambda_1\cdots a^\lambda_l\cdots b^\lambda_l \cdots
\alpha^\lambda_m \cdots \beta^\lambda_m \cdots]^T$. The transfer
matrices $[M^{\lambda}_{\lambda-1}]$ and $[M^{\lambda}_{\lambda+1}]$
are associated with the inner products appearing in
Eqs.(\ref{cs1}-\ref{cs4}). We ignore the details about their
definition. But it is not a difficult task to obtain them from the
above equations. The feature of the relation given in
Eq.(\ref{matrix}) is that the coefficient $[C^\lambda]$ is expressed
explicitly in terms of the two adjacent ones. If we want to work out
the next relation, i.e. to express $[C^{\lambda+1}]$ as a function
of $[C^\lambda]$ and $[C^{\lambda+2}]$, we need to use the above
procedure at the $(\lambda+1)$th segment. For example, to establish
a relation between $[C^{\lambda+1}]$ and $[C^\lambda]$ similar to
the ones given in Eqs.(\ref{cs3},\ref{cs4}), we need to use the
connection condition given by Eq.(\ref{continuityb}).
\par
What we must emphasize herein is that we can not obtain a simpler
transfer matrix to connect two sets of expansion coefficients
between the adjacent segments, namely, in the form of
$[C^\lambda]=[\mathcal {M}_{\lambda+1}^\lambda][C^{\lambda+1}]$.
This is due to that the numbers of modes in the neighbor segments
are different. Instead we can only obtain an iteration relation
involving three adjacent segments, as done above. Just due to such
an encumbrance, we can not express $[C^1]$ in terms of $[C^2]$
explicitly. However, by means of the wavefunction connection
conditions at the first junction interface, we can obtain the
following relation
\begin{equation}
[M^1_1][C^1]=[M^1_2][C^2]. \label{mfirst}
\end{equation}
Noting that the dimensions of the transfer matrices $[M^1_1]$ and
$[M^1_2]$ are, respectively, $J^1\times 2J^1$, and $J^1\times 2J^2$.
Both are not square matrices. By the same token, at the last
junction interface we obtain
\begin{equation}
[M^\sss{N}_\sss{N}][C^\sss{N}]=[M^\sss{N}_\sss{N-1}][C^\sss{N-1}]
\label{mlast}.
\end{equation}
From the above iteration relations, Eqs.(\ref{matrix}-\ref{mlast}),
and by a straightforward derivation we can eliminate all the
intermediate coefficients, as well as $[\alpha^1]$ and
$[\beta^\sss{N}]$. As a result, we obtain the scattering matrix
$[S]$ defined in Eq.(\ref{scattering}).

\section{Numerical results and discussions \label{result}}
After establishing the transfer matrix theory to describe the
electronic transport through the GMJ, we are now in a position to
perform the numerical calculation about the conductance spectrum in
some typical GMJ structures. At first, we consider a simplest case,
a graphene single junction formed by interconnecting the left and
right leads directly. We will focus mainly on the electronic
transport property in the vicinity of the Dirac point. Therefore, we
choose the two leads to be metallic ones which do not present any
finite bandgap between the conduction and valence bands. In such a
case the linear subband across the Dirac point provides a basic
channel for the electron transmission through the junction. The
calculated conductance spectra($\mathcal{G}$ vs $E$) for two single
junction structures are shown in Fig.2(a). We see that the spectrum
calculated by our transfer matrix approach agrees well with the one
obtained by the tight binding calculation in the low-energy region.
In particular, for the junction consisting of relatively wider
ribbons the agreement between the results of the two kinds of
methods is very satisfactory. Basically, this indicates that our
generalization of the transfer matrix method is successful to
describe the electron transport through the graphene junction.
\par
All these conductance spectra shown in Fig.2(a) exhibit the
staircase-like structures, which can be readily explained by the
match of the subband structures of the two metallic leads. An
interesting point to note in Fig.2(a) is that the conductance
spectrum shows a notable suppression in the vicinity of the Dirac
energy. In particular, a zero conductance occurs at the Dirac
energy. At first sight, such a conductance suppression is
inconsistent with the gapless subband structure of the two metallic
leads. The linear subbands of the two metallic leads always match
each other to provide an electron transmission mode. As a result,
one can expect that a nonzero conductance appears at the Dirac
point. But such a reasoning is dramatically in contradiction with
the calculated zero conductance. In fact, the nature of the
conductance suppression is the antiresonance effect, which we can
explain as follows. In the vicinity of the Dirac energy, only the
linear subbands are relevant to the electron transmission. Hence the
two metallic GNRs can be viewed as single mode quantum wires
coupling to each other directly. However, the zigzag-edged shoulder
induces a localized edge state with the eigen-energy equal to the
Dirac energy. Such a localized state couples to the linear subband
of the wider GNR. Consequently, when the electronic transport is
limited in the vicinity of the Dirac point, the graphene junction is
equivalent to the T-shaped quantum dot structure as shown in
Fig.2(b). The linear conductance of such a model has been
extensively studied,\cite{refA. Ueda,refY.Liu,refW.J.Gong} which can
be expressed in terms of the relevant parameters
\begin{equation}
\mathcal {G}(E)={e^2\over h}T(E)={e^2\over
h}\frac{4\xi}{(1+\xi)^2}\frac{[E-\varepsilon]^2}
{(E-\varepsilon)^2+[\frac{\Gamma}{2(1+\xi)}]^2},\label{CF}
\end{equation}
where $\Gamma=2\pi\rho_2 \tau^{2}$ and $\xi =\pi^2\rho_1\rho_2v^2 $
with $\rho_{1(2)}$ being the electron density of the states in two
leads. This expression presents a zero conductance at the quantum
dot level $\varepsilon$ indeed, which is called the antiresonance
effect. The antiresonance is in fact a destructive quantum
interference. The lateral quantum dot introduces new Feynman paths
with a phase shift $\pi$. As a result, the destructive quantum
interference occurs among electron Feynman paths. In the metallic
graphene junction, the edge state attached to the zigzag-edged
shoulder of the junction plays the role of the laterally coupled
quantum dot, which results in the antiresonance at the Dirac point.
In Fig.2(c) we calculate several conductance spectra when the width
of the narrower lead approaches the wider one whose width is fixed.
We find that the suppressed conductance valley becomes narrower as
the widths of the two leads approach each other. This effect can be
explained in terms of the above quantum dot model. One can readily
infer that the parameters $v$(the coupling between the two leads)
becomes larger, but $\tau$(the coupling between the propagating mode
in the wider lead and the edge state localized at the junction
shoulder) becomes smaller, as the two leads approach each other in
their widths. Consequently, from Eq.(\ref{CF}) one can deduce that
the conductance valley becomes narrower. On the contrary, when we
fix the width of the narrower lead, but to increase the width of the
wider lead, we can see from Fig.2(d) that conductance valley does
not vary notably. This is because that in such a situation the
parameter $\tau$, which mainly controls the valley width, does not
vary notably as the wider lead is further widened.
\par
In Fig.3(a) we show the calculated conductance spectrum for a
step-shaped GMJ. We take the case of $N=3$, namely, a two-junction
structure, slightly more complicated than a graphene single
junction. First, we can see that the results calculated by the tight
binding and transfer matrix methods, respectively, agrees with each
other very well. But the tight binding calculation is much more
time-consuming than the corresponding calculation via the transfer
matrix method. Secondly, we can see that the conductance valley
around the Dirac point found in a graphene single junction remains.
And it is further broadened.  In Fig.3(b), the width of the right
lead(the wider one) is fixed, the calculated spectrum shows that the
conductance valley gets narrower as the two shoulders becomes
shorter. In contrast, if we fix the width of the left lead,
meanwhile, to increase the width of the two shoulders, the
conductance valley is almost invariable. Such a case is shown in
Fig.3(c). These results shown in Fig.3(b) and (c) is consistent with
the size dependence of the graphene single junction, as shown in
Fig.2(c) and (d). This indicates that the conductance characteristic
in a single junction is preserved and further strengthened in the
two-junction structures. The results of the conductance spectra
shown in Fig.3(d) indicate that the profile of the steep conductance
valley does not vary sensitively with the increase of the
longitudinal size of the intermediate ribbon segment. This reflects
that the interaction between the two edge states at the two adjacent
shoulders is very weak.
\par
The conductance spectra for more complicated step-shaped and
T-shaped GMJs are shown in Fig.4(a) and (b), respectively. A common
feature of these spectra is that the conductance valley gets steeper
as the number of the junctions increases. As shown in Fig.4(a) and
(b), when $N> 3$ the profile of the conductance valley becomes
stable, unsensitive to the further increase of the number of the
junctions in the structures. This indicates that the conductance
valley in a single junction develops rapidly into a well-defined
insulating band in the GMJs. Following the antiresonance mechanism
mentioned above, it is not very difficult to understand the
appearance of the well-defined insulating band in the GMJ. As
claimed above, the graphene single junction in the low-energy
electron transport regime is equivalent to a lateral quantum dot
structure. In an analogous way a mapping between the GMJ and a
laterally coupled quantum dot chain, which is illustrated in
Fig.4(c), is reasonable. Of course, the edge states localized at the
individual shoulders of the junction interfaces play the roles of
the dangling quantum dots. And the quantum dots in the main chain
mimic the electron states in the ribbon segments. Each quantum dot
pair, made of the quantum dot in the main chain and its lateral
attachment, can be viewed as an antiresonance unit which contributes
a transmission probability amplitude, the squared modulus of which
has the same form as the transmission probability given by
Eq.(\ref{CF}). In the GMJ the electron wave will be reflected many
times by the multiple interfaces before it finally enters into the
right leads. The shortest transmission path is the one that the
electron penetrates all the junctions directly without any
reflection. Even in such a situation, the electron will undergo $N$
junctions, in other words, $N$ antiresonance units. Accordingly, the
corresponding transmission probability is proportional to the $N$th
power of $T(E)$, i.e. $~T(E)^\sss{N}$. ($T(E)$ is the transmission
probability of an individual antiresonance unit given by
Eq.(\ref{CF})). Thus the conductance suppression around the Dirac
point included in $T(\omega)$ of each quantum dot is efficiently
strengthened. As a result, the antiresonant valley in a single
junction evolves into a well-defined insulating band in a GMJ due to
multiple quantum interference. A comprehensive explanation about the
occurrence of the insulating band in a lateral quantum dot chain was
presented in the language of Feynman path in our previous
work,\cite{refW.J.Gong,refW.J.Gong2} which can be reasonably used to
account for the well-defined insulating band in the present GMJ
structures.
\par
In Fig.5 we show the conductance spectrum which is adjusted by
applying a gate voltage under some junctions interfaces. In our
theoretical treatment, the gate voltage is mimicked by introducing a
finite potential constant in the ribbon segment where the gate
voltage is applied. We can see that the insulating band is
effectively broadened when certain of the junction interfaces are
covered by the gate voltage. But the cost of such a broadening is
that the insulating band is not so well-defined than the one in the
absence of a gate voltage. Such a result can be readily understood
according to the multiple quantum interference picture explained
above. The transmission probability of each junction possesses a
valley center around the level of the localized edge state, in other
words, the quantum dot level in the corresponding quantum dot chain.
In the absence of a gate voltage such a level is just at the Dirac
point. On the other hand, in the presence of a gate voltage covering
partially some junctions of the total structure, the levels of the
edge states of these junctions, namely, the centers of these
antiresonance units, separates from those of other junctions free
from the gate voltage. Accordingly, the total transmission
probability presents a broadening of the insulating band, due to
that the gate voltage destroys the superposition of the
antiresonance centers of all antiresonance units. By the same
argument, one can readily understand that the insulating band only
exhibits a shift without an observable broadening when all junctions
are simultaneously tuned by a gate voltage. In light of the
sensitive adjustment of a gate voltage on the insulating band, as
shown in Fig.5, we suggest that the GMJ can be considered as a
device prototype of a nanoswitch.

\section{summary\label{summary}}
By generalizing the transfer matrix approach which was established
to describe the electron transport through quantum waveguide made of
conventional semiconductor materials, we provide an alternative
theoretical method to calculate the conductance spectrum of a GMJ,
in place of the tight-binding treatment. By comparing the calculated
results by the two methods, we find that our generalized transfer
matrix method works well in describing the electronic transport
properties of the GMJ structure if its size is not very small. In
particular, this method is especially suitable to deal with the GMJs
with relatively large sizes and many junctions. In such a case, it
is much more timesaving to calculate the conductance spectrum by
this method than by the tight-binding calculation. Interestingly,
with the transfer matrix method, we find that the GMJ exhibits a
well-defined insulating band around the Dirac point, which is due to
the strengthen of the intrinsic antiresonance effect of a graphene
single junction by the multiple quantum interference at many
junction interfaces. By virtue of the feature that the insulating
band can be sensitively tuned by a gate voltage, we propose to
design a nanoswitch based on this GMJ structure. Our generalized
transfer matrix method is also suitable to treat the graphene
waveguides with various junction shapes, such as the tortuous
nanoribbons with distinct bent angles.
\par
Finally, we need to make some remarks on the issue to implement the
function of nanoswitch, by means of the tunability of the insulating
band of the GMJ by a gate voltage. In fact, such a transport
phenomenon also appear in the T-shaped waveguides and the laterally
coupled quantum dot chains,\cite{refP.S.Deo,refW.J.Gong2} made of
the conventional semiconductor materials. The similar proposal to
make a nanoswitch was previously mentioned in these relevant work.
However, the occurrence of a well-defined insulating band in these
quantum structures requires that the fluctuation of the quantum dot
energy levels(the size of the stubs in a T-shaped waveguide) to be
sufficiently small. It is indeed a challenging task to fabricate
many identical quantum dots and connecting them periodically in a
circuit to realize the function of nanoswitch. However, in the
present GMJ structures, the problem of energy level fluctuation is
automatically avoided since the antiresonant levels are provided by
the localized edge state at the zigzag edge, rather than the quantum
confinement in the stub region. The energy levels of these localized
states are automatically aligned with  each other at the Dirac
point, regardless of the details of the shape of the GMJ. This
implies the prefect realization of the uniform antiresonance levels.
Generally, the insulating band can be easily formed in the
step-shaped as well as the T-shaped GMJs. It does not rigidly
require the identity of the shapes of every junctions. In addition,
it is experimentally demonstrated that ballistic transport can be
retained in graphene nanostructures over sub-micron scale. This
means that the quantum interference can play a dominate role even
the size of the GMJ becomes very large. Thus the insulating band
structure is expected to remain in a GMJ even with a size up to
sub-micron scale. To sum up these feature, we can conclude that the
GMJs are the optimal candidate as the prototype of the nanoswitch.

 This work was financially supported by the National Nature Science
Foundation of China under Grant NNSFC10774055, and the specialized
Research Fund for the Doctoral Program of Higher
Education(SRFDP20070183130).

\renewcommand{\baselinestretch}{1}

\clearpage

\begin{figure}
\caption{(Color online) (a) Schematic illustration of a graphene
multiple junction structure. (b) The honeycomb lattice of an
intermediate part of the graphene multiple junction, around the
$\lambda$th junction. The carbon atoms belonging to the two distinct
sublattices, A and B, are distinguishingly labeled as A:$\bullet$
and B:$\circ$. The atoms at the edges linked to the interior ones by
the dashed lines stand for the truncated atoms to shape the
GMJ.\label{schematic}}
\end{figure}

\begin{figure}
\caption{(Color online) (a) The conductance spectrum of two graphene
single junctions with distinct sizes. Comparison is made between the
conductance spectra calculated by the tight binding(TB) and the
transfer matrix(TM) methods. (b) Schematic of a laterally coupled
quantum dot structure, where $v$ and $\tau$ represent, respectively,
the coupling coefficients between the two leads and between the
quantum dot and the right lead. The calculated conductance spectrum
from this model is also plotted. (c) A comparison of the conductance
spectra of the graphene single junctions by changing the width of
the left lead, whereas the width of the right lead is fixed. (d)
Similar to (c), but the width of the right lead is changed with the
left lead fixed. \label{conductance1}}
\end{figure}

\begin{figure}
\caption{(Color online) Conductance spectrum of step-shaped
two-junction structures(GMJs of \emph{N}=3). (a) A comparison of the
calculated results by the tight binding and the transfer matrix
methods. (b) and (c) A comparison of the conductance spectrum by
changing the width of the junction shoulders. The two different
cases with the width of the left and right leads fixed,
respectively, are shown in (b) and (c). (d) A comparison of the
conductance spectrum by changing the longitudinal size $L_{2}$ of
the intermediate segment.
 \label{conductance2}}
\end{figure}

\begin{figure}
\caption{(Color online) Conductance spectrum of typical GMJ
structures. The length of the intermediate ribbon segment is
$L_{\lambda}=48/\sqrt{3}$. (a) The result of the step-shaped
structures. (b) The result of the T-shaped structures. (c) A
illustration of the laterally coupled quantum dot chain, which is
used to account for the insulating band in the conductance spectrum
of the GMJ. \label{conductance3}}
\end{figure}

\begin{figure}
\caption{(Color online) Conductance spectrum of GMJ structures in
the presence of the gate voltage. (a) The result of the step-shaped
structures. (b) The result of the T-shaped structures. $V_\lambda$
denotes the potential constant produced by the gate voltage in the
$\lambda$ segment. \label{conductance4}}
\end{figure}

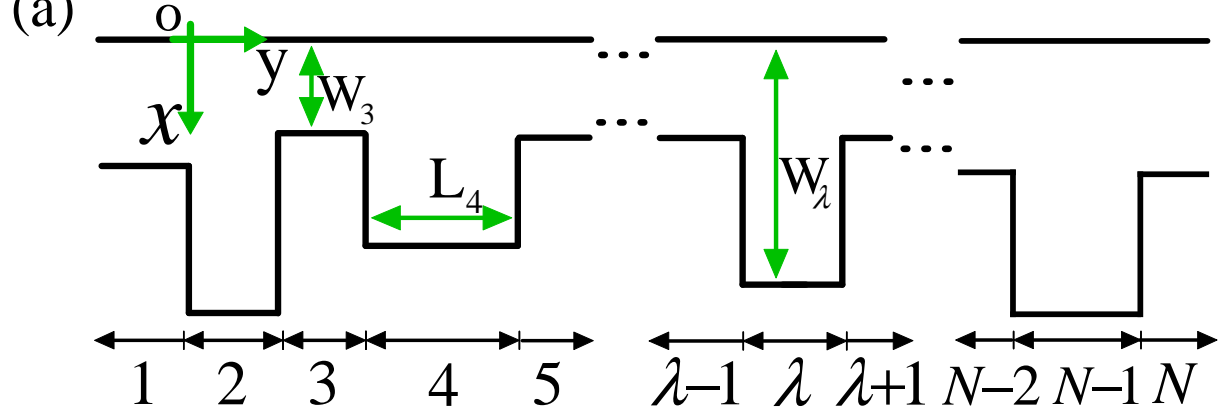
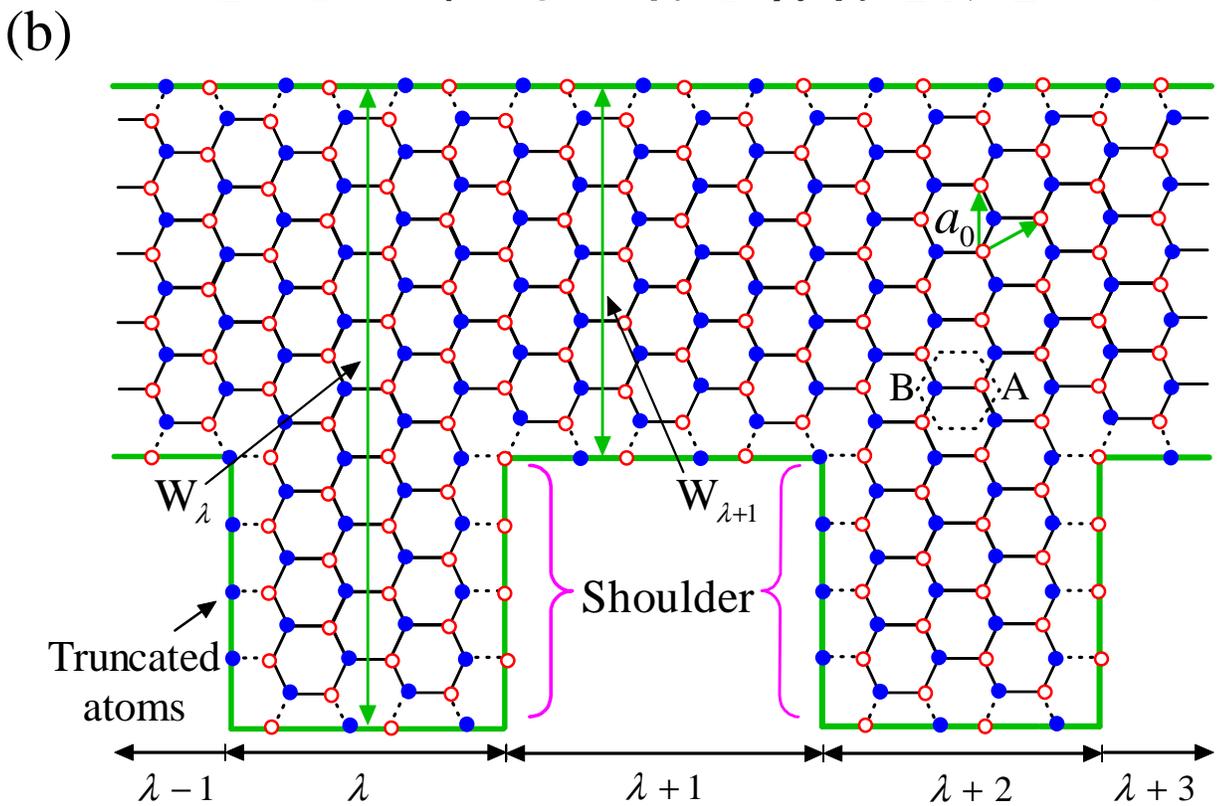
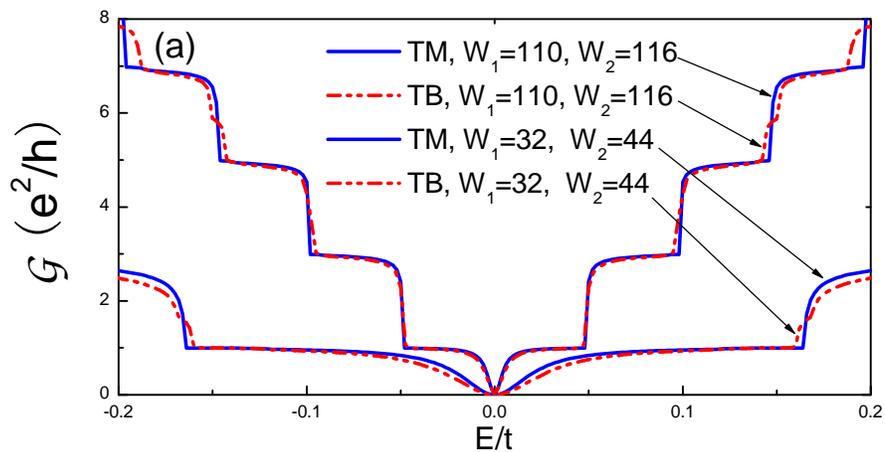
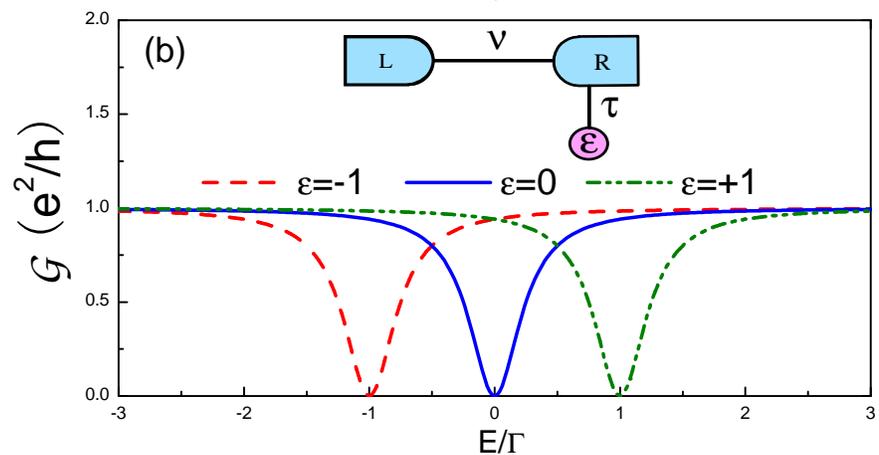
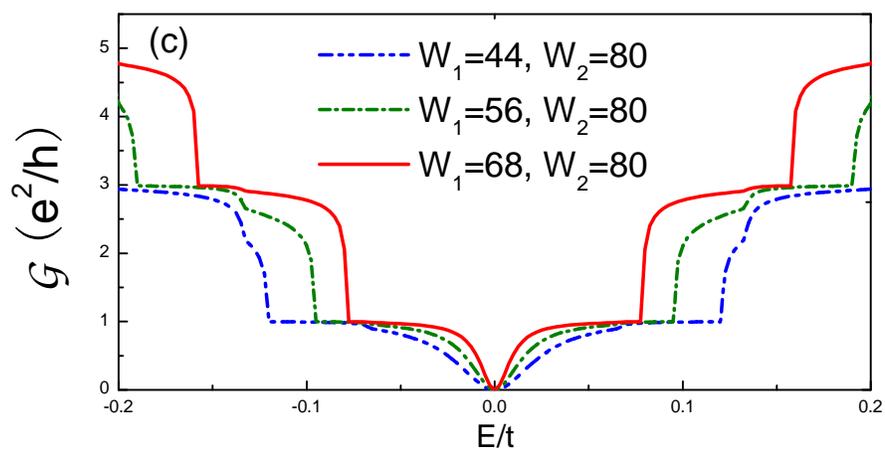
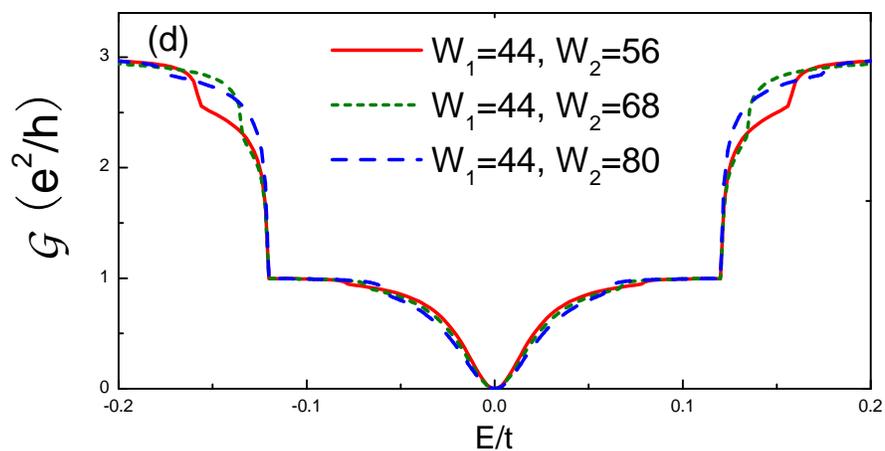
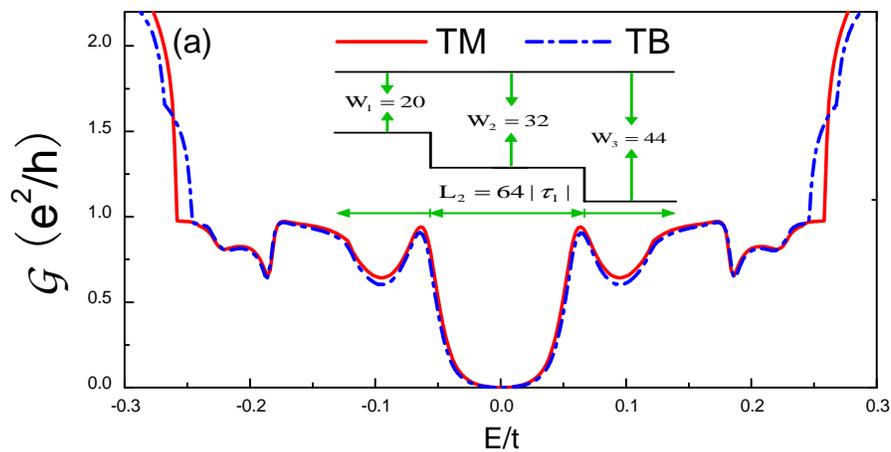
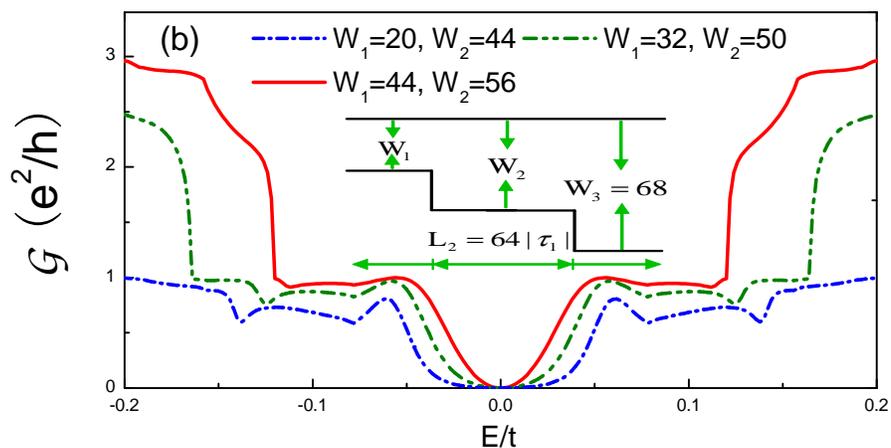
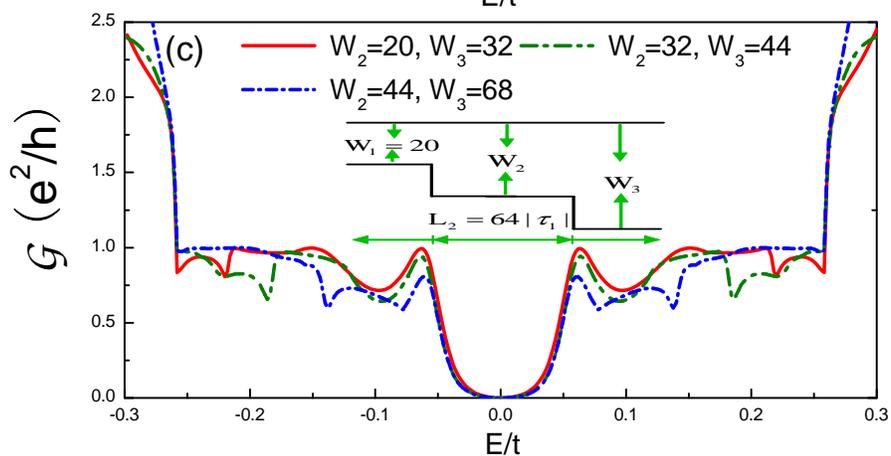
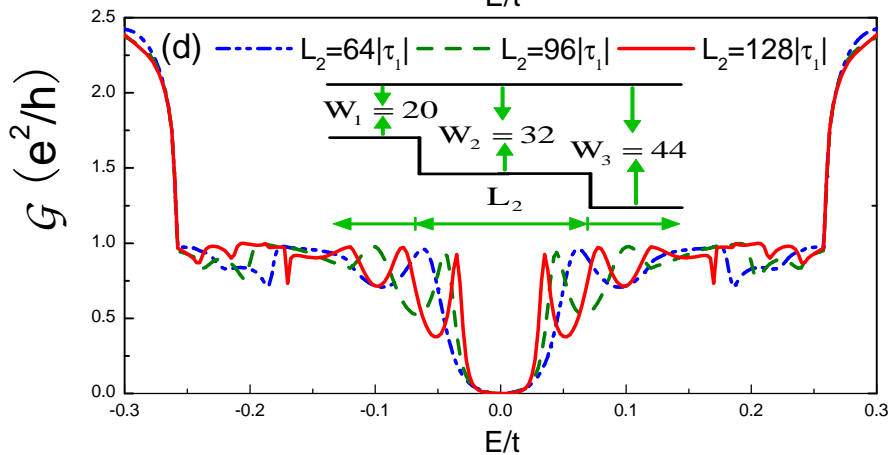
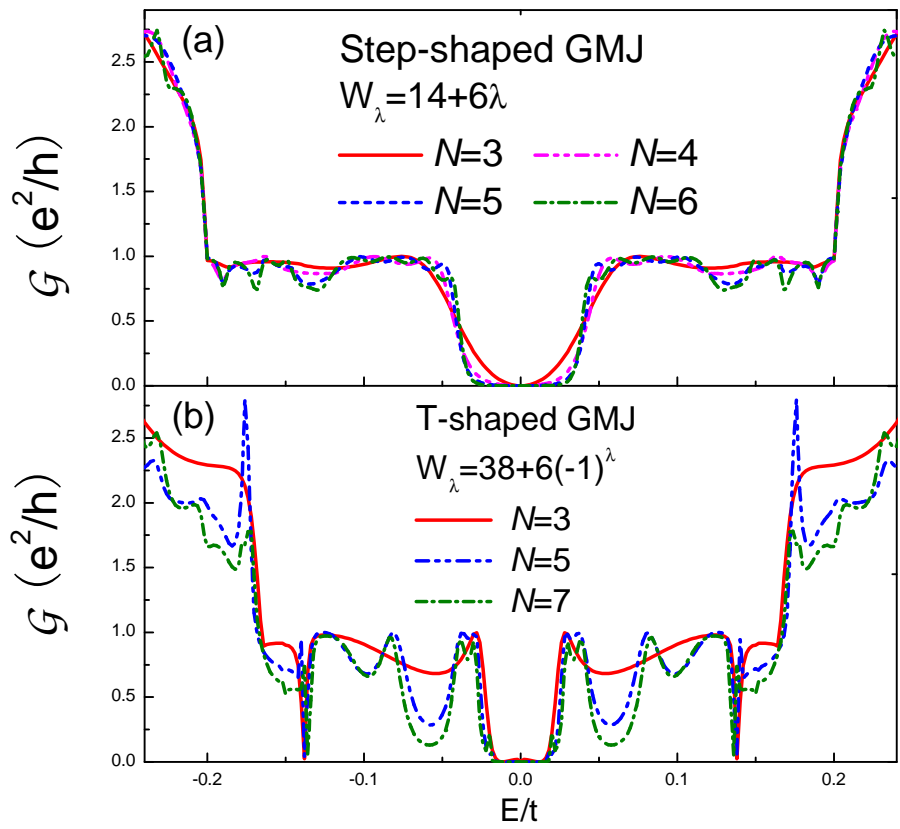
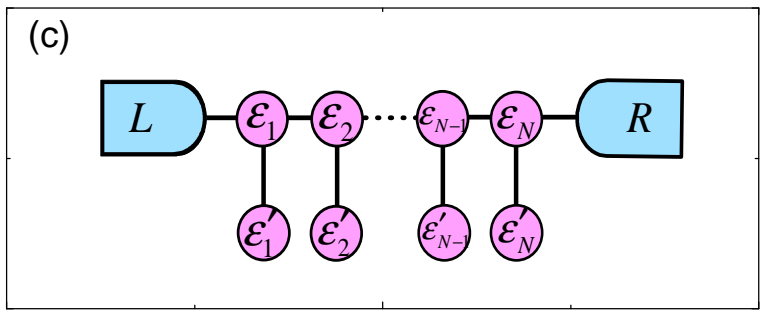
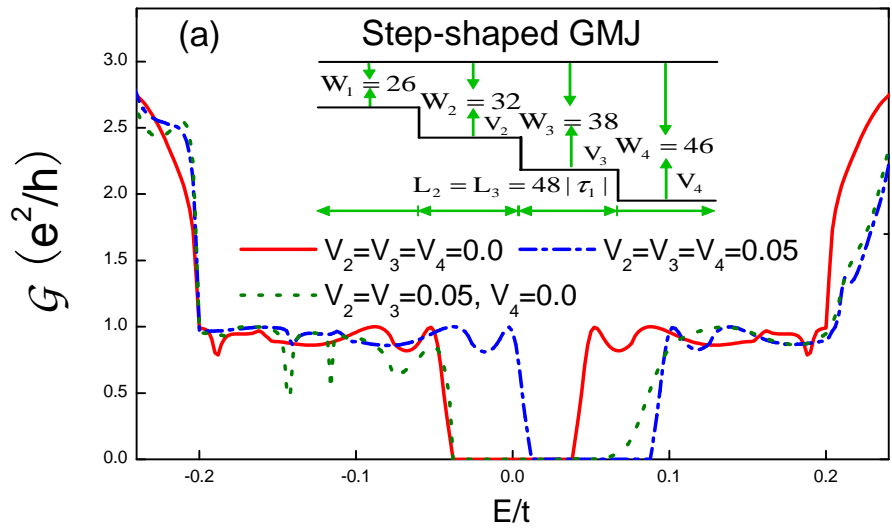
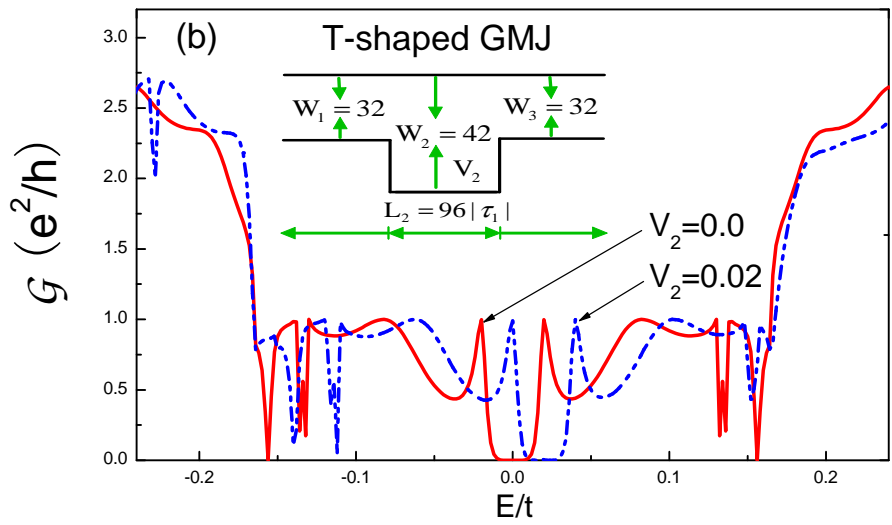
\end{document}